\documentclass[a4]{elsart}

 \usepackage{graphicx}
\usepackage{amssymb}

\begin{document}

\bibliographystyle{elsart-num}
\begin{frontmatter}


\title{ESR of YbRh$_{2}$Si$_{2}$ and $^{174}$YbRh$_{2}$Si$_{2}$ : local and itinerant properties}
 \author[MPI]{J. Wykhoff},  
 \author[MPI]{J. Sichelschmidt\corauthref{cor1}}, \ead{Sichelschmidt@cpfs.mpg.de}
 \author[CEA]{G. Lapertot},       
 \author[CEA]{G. Knebel}, 
 \author[CEA]{J. Flouquet},
 \author[Kazan]{I.I. Fazlishanov},
 \author[EKM]{H.-A. Krug von Nidda},
 \author[MPI]{C. Krellner},
 \author[MPI]{C. Geibel},
 \author[MPI]{F. Steglich}
%\thanks[label2]{a}


\address[MPI]{Max-Planck-Institut f\"ur Chemische Physik fester Stoffe, 01187 Dresden, Germany}
\address[CEA]{Departement de la Recherche Fondamentale sur la Matiere Condensee, SPSMS, CEA Grenoble, 38054 Grenoble, France}
\address[EKM]{Experimentalphysik V, EKM, Universit\"at Augsburg, 86135 Augsburg, Germany}
\address[Kazan]{E. K. Zavoisky Physical-Technical Institute, 420049 Kazan, Russia}

\begin{abstract}
Below the Kondo temperature the heavy Fermion compound YbRh$_{2}$Si$_{2}$ shows a well defined Electron Spin Resonance (ESR) with local Yb$^{3+}$ properties. We report a detailed analysis of the ESR intensity which gives information on the number of ESR active centers relative to the ESR of well localized Yb$^{3+}$ in YPd$_3$:Yb. The ESR lineshape is investigated regarding contributions from itinerant centers. From the ESR of monoisotopic $^{174}$YbRh$_{2}$Si$_{2}$ we could exclude unresolved hyperfine contributions to the lineshape.
\end{abstract}

\begin{keyword}
Electron spin resonance \sep heavy fermion \sep YbRh$_{2}$Si$_{2}$ 
\PACS 75.20.Hr \sep 76.30.Kg
\end{keyword}

\corauth[cor1]{Corresponding author.}

\end{frontmatter}
%\begin{linenumbers}
%%%%%%%%%%%%%%%%%%%%%%%%%%%%%%%%%%%%%%%%%%%%%%%%%%%%%%%%%%%%%%%
\section{Introduction}
The heavy fermion compound YbRh$_{2}$Si$_{2}$ shows a well defined Electron Spin Resonance (ESR) signal which features properties similarly observed for local Yb$^{3+}$ ions in metallic environments \cite{sic03}. Remarkably, this signal was observed at resonance energies much smaller than the Kondo energy and well \textit{below} the Kondo temperature $T_{\rm K}\approx 25$~K, where the magnetic Yb$^{3+}$ moments are supposed to be largely screened and ESR silent. 
%This seeming contradiction poses an important, yet open question to the nature of magnetism in YbRh$_{2}$Si$_{2}$.
This seemingly contradiction is an important, yet unexplained feature of the magnetism in YbRh$_{2}$Si$_{2}$. This compound is located very close to a magnetic field induced quantum critical point where weak antiferromagnetic (AF) ordering is suppressed in a field of 70~mT applied in the magnetic easy tetragonal $ab$ plane \cite{Tro00}. Besides the AF order at $T_{\rm N}=70$~mK pronounced ferromagnetic fluctuations are evident from a highly enhanced Sommerfeld-Wilson ratio \cite{Geg05} and from the spin dynamics as seen in $^{29}$Si NMR measurements \cite{Ish02}.\\
The existence of the ESR in YbRh$_{2}$Si$_{2}$ below $T_{\rm K}$ seems even more surprising when its local character is envisaged: besides its Weiss-like temperature dependence of the intensity it shows a pronounced anisotropy which perfectly agrees with the magnetocrystalline anisotropy \cite{sic07}. However, results of $^{170}$Yb M\"ossbauer experiments provided the most direct evidence that the observed ESR line is much to narrow to be assigned to local Yb$^{3+}$ spin dynamics \cite{kne06}. On the other hand, assuming the local moments to be Kondo screened, conduction carriers themselves could cause a narrow and intense ESR signal because strong ferromagnetic fluctuations are present. This kind of conduction electron spin resonance is reported, for instance, in Pd \cite{mon78} or in TiBe$_{2}$ \cite{sha87}.\\ 
With this paper we highlight how both scenarios (local / itinerant) characterize the ESR spectra. For this purpose we present (i) an estimation of the ESR intensity which is involved in the YbRh$_{2}$Si$_{2}$ ESR compared to the local Yb$^{3+}$ ESR of YPd3:Yb and (ii) a discussion of the lineshape's itinerant character which may be seen if spin diffusion is relevant in the relaxation process. For the latter we also investigated the ESR of mono-isotopic $^{174}$YbRh$_{2}$Si$_{2}$ in order to take into account eventual contributions from unresolved hyperfine lines.
%
%%%%%%%%%%%%%%%%%%%%%%%%%%%%%%%%%%%%%%%%%%%%%%%%%%%%%%%%%%%%%%%
\section{Experimental details}
ESR probes the absorbed power $P$ of a transversal magnetic microwave field as a function of an external, static magnetic field $B$. To improve the signal-to-noise ratio, a lock-in technique is used by modulating the static field, which yields the derivative of the resonance signal $dP/dB$. The ESR experiments were performed at X--band frequencies ($\nu\,\approx\,9.4$~GHz) with a BRUKER ELEXSYS 500 spectrometer. The temperature was varied between 2.7~K$\leq T\leq$ 300~K by using a He-flow cryostat. For the ESR-measurements down to the lowest accessible temperature of $T=0.69$~K a $^3$He cold-finger bath cryostat was used.\\
We used high quality single crystalline platelets of YbRh$_{2}$Si$_{2}$ and $^{174}$YbRh$_{2}$Si$_{2}$ with small residual resistivities as low as $\rho_0=0.5\,\mu\Omega\rm{cm}$ and with very sharp anomalies in the specific heat at $T=T_N$. Their preparation as well as their magnetic and transport properties were thoroughly described: YbRh$_{2}$Si$_{2}$ in refs. \cite{Tro00,Geg02} and $^{174}$YbRh$_{2}$Si$_{2}$ in refs. \cite{kne05,kne06}. The crystals were mounted in the microwave cavity such that the microwave magnetic field was always within the tetragonal basal plane.\\
The polycrystalline sample of YPd$_3$:Yb was prepared by argon-arc-melting stoichiometric amounts of Y, Pd and dopant amounts of YbPd$_3$ \cite{gam83}. The latter was prepared before in an induction furnace because the usage of pure Yb metal would lead to a considerable Yb loss due to its high vapour pressure. Debye-Scherrer X-ray diffraction confirmed a single-phase sample. A Yb concentration of 0.6\% was determined by SQUID magnetization measurements.      
%
%%%%%%%%%%%%%%%%%%%%%%%%%%%%%%%%%%%%%%%%%%%%%%%%%%%%%%%%%%%%%%%
\section{Results and Discussion}
\begin{figure}[ht]
	\includegraphics*[width=8cm]{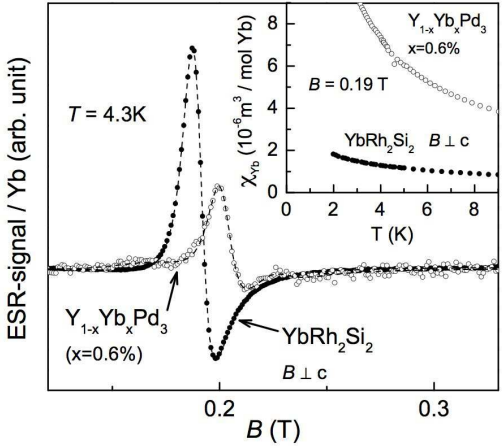}
	\caption{ESR signals (${\rm d}P/{\rm d}B$) of YbRh$_{2}$Si$_{2}$ (single crystal, $B\perp c$) and Yb$^{3+}$ in YPd$_3$:Yb (polycrystal) at 9.4~GHz and $T=4.3$~K normalized to the Yb content. Dashed lines denote fitcurves: one Lorentzian shape for YbRh$_{2}$Si$_{2}$ and three Lorentzians for YPd$_3$:Yb in order to account for the weak hyperfine structure wings. The YbRh$_{2}$Si$_{2}$ data correspond to a sample batch with the lowest residual ESR linewidth. Inset: magnetic susceptibility $\chi_{\rm Yb}$ normalized to the Yb content and measured at the ESR resonance field $B\,=\,0.19$~T.}
	\label{Fig1}
\end{figure}

We first address the question how much intensity is involved in the YbRh$_{2}$Si$_{2}$ - ESR compared to the ESR of a well localized Yb ion, Y$_{\rm1-x}$Yb$_{\rm x}$Pd$_3$, x=0.6\%. The latter compound is suitable for this comparison because it displays ESR properties typical of a local Yb$^{3+}$ magnetic moment. The ESR linewidth is well characterized by a Korringa (temperature linear) relaxation and a first excited ($\Gamma_8$) level at $\approx50$~K. The temperature independent effective $g$-factor of 3.34 is compatible with a doublet $\Gamma_7$ groundstate \cite{gam83,ret81}.\\ 
The main frame of Fig. \ref{Fig1} shows the ESR signals of both above mentioned compounds, recorded at $T\,=\,4.3\,$K with the same experimental conditions. Note that Fig. \ref{Fig1} shows the ESR signal amplitudes after normalization to the amount of Yb ions by proper consideration of the molar volume which is probed by the microwave in the skin depth ($\approx 1.3\,\mu$m for YbRh$_{2}$Si$_{2}$, $\approx 2.3\,\mu$m for Y$_{\rm1-x}$Yb$_{\rm x}$Pd$_3$).
From simply inspecting Fig. \ref{Fig1} it is already clear at first sight that the ESR of YbRh$_{2}$Si$_{2}$ cannot be caused by a few percent of local Yb$^{3+}$ moments. It is even obvious that the previously assessed lower bound of 60\% ESR active Yb ions \cite{sic04} is underestimated. With respect to a more accurate statement we compare in the following the bulk magnetic susceptibility with the ESR intensity which is proportional to the spin susceptibility of the ESR probe. In general, the ESR intensity can be determined by the area $\propto Amp \cdot \Delta B^2$ under the ESR absorption signal. We determined the parameters $Amp$ (signal amplitude) and $\Delta B$ (linewidth) by fitting the spectra with metallic Lorentzian line shapes which take the skin effect into account. The hyperfine structures which are visible in the wings of the Y$_{\rm1-x}$Yb$_{\rm x}$Pd$_3$, ${\rm x}\,=\,0.6$\% spectra \cite{gam83} were considered with two additional Lorentzians. Using the Yb-normalized spectra as shown in Fig. \ref{Fig1} we found that the ESR intensity of Y$_{\rm1-x}$Yb$_{\rm x}$Pd$_3$, x=0.6\% \textit{exceeds} the ESR intensity of YbRh$_{2}$Si$_{2}$ by a factor of $1.5\pm0.2$. However, as can be seen in the inset of Fig. \ref{Fig1}, the bulk magnetic susceptibility per Yb, $\chi_{\rm Yb}$, of Y$_{\rm1-x}$Yb$_{\rm x}$Pd$_3$, x=0.6\% also \textit{exceeds} $\chi_{\rm Yb}$ of YbRh$_{2}$Si$_{2}$, in fact, by even a larger factor. We suspect that some of the discrepancy in the factors is due to uncertainties in the estimation of hyperfine contributions in Y$_{\rm1-x}$Yb$_{\rm x}$Pd$_3$, x=0.6\%. Albeit these uncertainties our comparison with a local Yb$^{3+}$ ESR probe shows that \textit{all} of the Yb ions in YbRh$_{2}$Si$_{2}$ contribute to the observed ESR signal.\\  
Next, we check the assumption that unscreened, local Yb$^{3+}$ moments could cause an ESR signal with a characteristic energy which is much smaller than the Kondo temperature. In this case a hyperfine coupling between the 4$f$ electron and the Yb nuclear spin should contribute to the ESR spectra.  $^{171}$Yb and $^{173}$Yb nuclei may provide hyperfine energies near 1~K \cite{kne06}. We therefore investigated the ESR of monoisotopic $^{174}$YbRh$_{2}$Si$_{2}$.  The isotope $^{174}$Yb has zero nuclear spin and hence no hyperfine interactions could be involved in a putative Yb$^{3+}$ ESR in $^{174}$YbRh$_{2}$Si$_{2}$. 
\begin{figure}[ht]
	\includegraphics*[width=8cm]{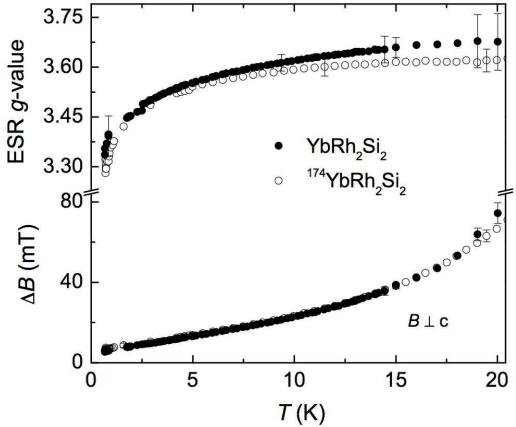}
	\caption{The ESR $g$-value and linewidth for YbRh$_{2}$Si$_{2}$ and $^{174}$YbRh$_{2}$Si$_{2}$ single crystals ($B\perp c$) at 9.4~GHz (X--band). The depicted data correspond to two samples with a comparable residual ESR linewidth. Note that both data-sets nicely coincide, i.e. hyperfine interactions are not relevant for the ESR parameters.}
	\label{Fig2}
\end{figure}
The temperature behavior of ESR $g$-value and linewidth $\Delta B$ of mono-isotopic and mixed-isotopic YbRh$_{2}$Si$_{2}$ is shown in Fig. \ref{Fig2} and the corresponding ESR spectra are plotted in Fig. \ref{Fig3}b for a representative temperature. $\Delta B(T)$ reflects a typical ESR behavior of a magnetic moment in a metallic host \cite{sic03,wyk07}. Both data-sets in Fig. \ref{Fig2}, $g$-value and linewidth $\Delta B$, coincide well within the experimental error in the complete accessed temperature range. Therefore, hyperfine interactions obviously do not contribute to the ESR relaxation and resonance field in YbRh$_{2}$Si$_{2}$. As will be shown below, the lineshape yields also no indications for hyperfine contributions.
We point out that for the data in Fig. \ref{Fig2} we used two samples with similar linewidth and similar residual resistivity ratio. As reported earlier \cite{wyk07} the linewidth at a given temperature is related to the residual resistivity ratio which in turn determines the residual linewidth $\Delta B_0$ (temperature-linear part of $\Delta B$ extrapolated to zero temperature). The temperature dependence of the linewidth obeys the scaling $\Delta B^* = (\Delta B -\Delta B_0)/\Delta B_0$, i.e. $\Delta B^*(T)$ for all investigated YbRh$_{2}$Si$_{2}$, for La-doped YbRh$_{2}$Si$_{2}$ \cite{wyk07} and also for $^{174}$YbRh$_{2}$Si$_{2}$ collapse onto one single curve. This scaling provides evidence that the lattice relaxation of a strongly coupled Yb$^{3+}$ -- conduction electron system is an important ingredient for understanding the ESR linewidth in YbRh$_{2}$Si$_{2}$. This situation resembles the bottleneck-relaxation mechanism which was discussed extensively for diluted magnetic moments in metallic hosts \cite{bar81}.\\
From a putative ESR of a strongly coupled Yb$^{3+}$ -- conduction electron system one would expect typical features of an itinerant ESR probe. For example the lineshape should be influenced by the ESR probe spin diffusion as theoretically described by {F.J.}~Dyson \cite{dys55} and experimentally verified by G.~Feher and {A.F.}~Kip \cite{feh55}. Applying Dyson's lineshape, we found that the spectra cannot be described better when using a finite spin diffusion time. Hence, the observed lineshape seems to agree well with the limit of infinite spin diffusion in Dyson's formalism which corresponds to the usual metallic Lorentzian lineshape of local moments in metals. However, as shown in Fig. \ref{Fig3}a, a close-up view of the ESR line flanks reveals systematic differences between a metallic Lorentzian fit and the data. These deviations could not simply be explained by a superposition of two metallic Lorentzian lines, for instance, and they are found for YbRh$_{2}$Si$_{2}$ samples of 5 investigated batches and for the $^{174}$YbRh$_{2}$Si$_{2}$ sample as well. 
\begin{figure}[ht]
	\includegraphics*[width=8cm]{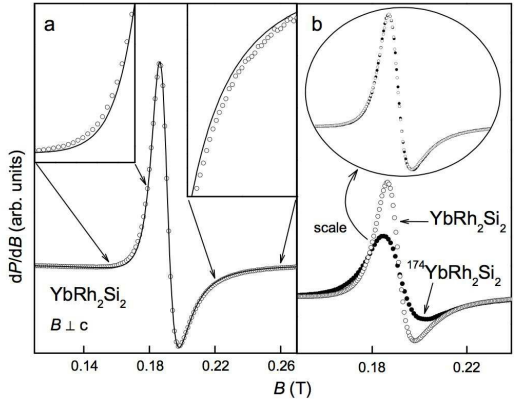}
	\caption{Left frame: YbRh$_{2}$Si$_{2}$ ESR-line (circles) at $T = 5$~K and metallic Lorentzian fit (solid line). Inserts show enlarged view of the line wings where the fit systematically deviates from the data. Right frame: Comparison of the ESR-lines of mono-isotopic (zero Yb nuclear spin) and mixed-isotopic YbRh$_{2}$Si$_{2}$ at $T\,=\,5$~K (sample batch with the lowest residual ESR linewidth was used). Although the particular parameters of the lines are different their lineshapes agree very well since both spectra can be scaled on top of each other as described in the text. This demonstrates that no hyperfine structures due to Yb nuclear spins contribute to the lineshape.}
	\label{Fig3}
\end{figure}
We also checked whether the lineshape distortion arises from unresolved hyperfine line-splitting in the ESR of YbRh$_{2}$Si$_{2}$ with mixed Yb isotopes. A hyperfine line-splitting is expected for magnetic ions with nonzero nuclear spin moment and, for instance, is reported for the isotope $^{171}$Yb diluted in a Au metallic matrix \cite{spa83}. In dense systems like YbRh$_{2}$Si$_{2}$ such line splittings might be unresolved due to a strong exchange narrowing \cite{bar81} but, however, could cause the ESR lineshape to deviate from a Lorentzian form. In Fig.~\ref{Fig3}b we compare the ESR lineshape of YbRh$_{2}$Si$_{2}$ (nonzero Yb nuclear spin) and $^{174}$YbRh$_{2}$Si$_{2}$ (zero nuclear spin). We scaled both spectra by stretching and shifting the abscissa and ordinate-values so that they fall on top of each other. The circled insert of Fig.~\ref{Fig3}b shows the result of this scaling which demonstrates that unresolved hyperfine structures do not contribute to the ESR lineshape of YbRh$_{2}$Si$_{2}$.  
%
%%%%%%%%%%%%%%%%%%%%%%%%%%%%%%%%%%%%%%%%%%%%%%%%%%%%%%%%%%%%%%%
\section{Conclusion}
We compared the ESR of YbRh$_{2}$Si$_{2}$ with the ESR of a system which contains a well defined, local Yb$^{3+}$ moment, Y$_{\rm1-x}$Yb$_{\rm x}$Pd$_3$, x=0.6\%. From relating the ESR intensities of these compounds to their Yb content and taking into account their difference in magnetic susceptibility we conclude that the ESR of YbRh$_{2}$Si$_{2}$ involves the magnetic moments of \textit{all} its Yb ions.\\ 
Assuming an itinerant ESR probe in YbRh$_{2}$Si$_{2}$ a corresponding lineshape description within the theory of F.J. Dyson \cite{dys55} yields a spin diffusion time in the infinite limit. However, such a description still leaves small systematic differences between lineshape fit and data which cannot be explained by unresolved hyperfine structures. We suspect that these differences are related to the ESR spin dynamics of the strongly coupled Yb$^{3+}$ -- conduction electron system in YbRh$_{2}$Si$_{2}$.    \\

%
%%%%%%%%%%%%%%%%%%%%%%%%%%%%%%%%%%%%%%%%%%%%%%%%%%%%%%%%%%%%%%%
%
{\bf Acknowledgement}\\
The low-temperature ESR-measurements at the University of Augsburg were partially supported by the German BMBF under Contract No. VDI/EKM13N6917 and by the Deutsche Forschungsgemeinschaft within SFB484 (Augsburg). I.I. Fazlishanov acknowledges support by the Volkswagen Foundation (I/82203).
%
%\end{linenumbers}
%

\end{document}